\documentclass[11pt]{article}
\usepackage[centertags]{amsmath}
\usepackage{amssymb}
\usepackage{graphicx}

\newtheorem{thm}{Theorem}[section]
\newtheorem{defn}[thm]{Definition}
\newtheorem{cor}[thm]{Corollary}
\newtheorem{lem}[thm]{Lemma}
\newtheorem{prop}[thm]{Proposition}
\newtheorem{rem}[thm]{Remark}

\def\qed{\hfill {\vrule height5pt width5pt depth2pt}}
\begin{document}

\title{\bf Synchronized Dynamics and Nonequilibrium
Steady States in a Stochastic Yeast Cell-Cycle Network}

\author{Hao Ge\footnote{
School of Mathematical Sciences, Peking University, Beijing
100871, P.R.C.; email: edmund\_ge@tom.com} \and Hong
Qian\footnote{Department of Applied Mathematics, University of
Washington, Seattle, Washington 98195, U.S.A.; email: qian@amath.washington.edu}
\and Min Qian\footnote{School of Mathematical Sciences, Peking University,
Beijing 100871, P.R.C.}
       }
\maketitle{}
\date{}

\begin{abstract}
Applying the mathematical circulation theory of Markov chains, we
investigate the synchronized stochastic dynamics of a discrete
network model of yeast cell-cycle regulation where stochasticity has
been kept rather than being averaged out. By comparing the network
dynamics of the stochastic model with its corresponding
deterministic network counterpart, we show that the synchronized
dynamics can be soundly characterized by a dominant circulation in
the stochastic model, which is the natural generalization of the
deterministic limit cycle in the deterministic system. Moreover, the
period of the main peak in the power spectrum, which is in common
use to characterize the synchronized dynamics, perfectly corresponds
to the number of states in the main cycle with dominant circulation.
Such a large separation in the magnitude of the circulations,
between a dominant, main cycle and the rest, gives rise to the
stochastic synchronization phenomenon.
\begin{flushleft}
{\bf KEY WORDS:} Boltzmann machine; circulation theory; Hopfield
network; nonequilibrium steady state; synchronization; yeast cell
cycle
\end{flushleft}
\end{abstract}

\section{Introduction}

    Synchronization is an important characteristics of many biological
networks \cite{sync,winfree} whose dynamics has been modelled
traditionally by deterministic, coupled nonlinear ordinary
differential equations in terms of regulatory mechanisms and kinetic
parameters \cite{Mu,FMWT}. Two important classes of biological
networks which have attracted wide attentions in recent years are
neural networks \cite{Scott} and cellular biochemical networks
\cite{Goldbeter}. A deterministic model, however, only describes the
averaged behavior of a system based on large populations; it can not
capture the temporal fluctuations of a small biological system with
either extrinsic or intrinsic noise. For example, neuronal firing
has an inherent variability, and many biochemical regulatory
networks inside a cell consist of molecular species with low
concentrations, where stochastic models with chemical master
equations (CME) based on biochemical reaction stoichiometry,
molecular numbers, and kinetic rate constants should be developed.
Such an approach has already provided important insights and
quantitative characterizations of a wide range of biochemical
systems \cite{Mc,Van,Fox1,Fox2,QH1,Zhou}. We refer to \cite{WD} for
a good introduction to stochastic modelling in biology.

    As there is a growing awareness and interest in studying
the effects of noise in biological networks, it becomes more and
more important to quantitatively characterize the synchronized
dynamics mathematically in stochastic models, because the concepts
of limit cycle and fixed phase difference no longer holds in this
case. Instead, physicists and biologists always have to characterize
synchronized dynamics by the distinct peak of its power spectrum or
just only by observing the stochastic trajectories, which however
may cause ambiguities in the conclusion. Therefore, a logical
generalization of limit cycle in stochastic models needs to be
developed.

    On the other hand, from the view of statistical physics, these stochastic models for
systems cell biology exhibit nonequilibrium steady states (NESS) in
which nonequilibrium cycle fluxes necessarily emerge
\cite{hqjpcm05}. The relation between an NESS and traditional
nonlinear dynamics, in particular bistability \cite{hqprl05} and
limit-cycle oscillations, has been discussed recently
\cite{hqjpc06,QH1}.

    We have recently developed a rather complete mathematical theory of
nonequilibrium steady state (NESS, \cite{JQQ}) of open systems,
taking Markov chains as the model of biochemical systems. One of the
most important concepts in the mathematical theory of NESS is the
circulation(also called cycle fluxes) \cite{JQQ}, which corresponds
to the cycle kinetics in open chemical systems \cite{hill,hqjpcm05}.
Actually, since the stochastic circulation in NESS is defined as the
time-averaged frequency of each cycle in the sense of trajectory, it
can also be regarded as a quantitative characterization for the
intensities of stochastic synchronized dynamics in different
attractor basins. This paper will thoroughly investigate the
interplay described above in a currently interested stochastic
cell-cycle network model.

Cell cycle is one of important biological processes whose underlying
molecular networks have been extensively studied. Tyson and his
coworkers have studied mathematical models for yeast cell cycle
based on differential equations \cite{tyson1}. Their key finding is
that cell cycle is a hysteresis loop, via saddle-node bifurcations,
driven by the periodic changes in cell mass due to cell growth and
division \cite{FMWT}.  The biological ``check points'' correspond to
the steady states of the dynamical system \cite{tyson2}.

\begin{figure}[h]
\centerline{\includegraphics[width=2.5in,height=3in]{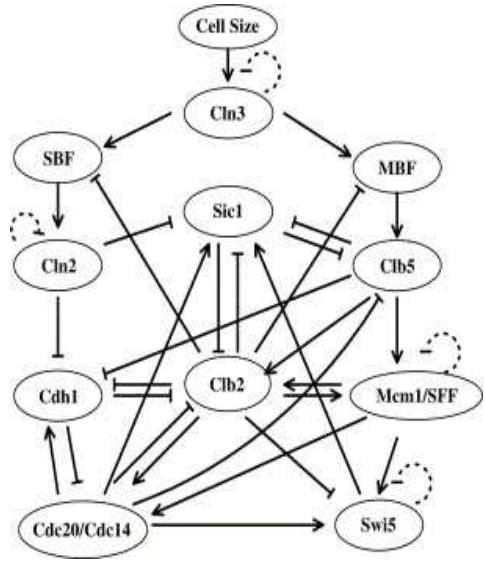}}
\caption[fig_diagram]{The cell-cycle network of the budding yeast.
Each node represents a protein or a protein complex. Arrows are
positive regulation, ¡°T¡±-lines are negative regulation, dotted
¡°T¡±-loops are degradation. It is just Fig. 1 from \cite{ZQ1} with
permission.} \label{fig_diagram}
\end{figure}

However, in many cellular biochemical modelling, a detailed,
molecular number-based CME model is not warranted because of a lack
of quantitative experimental data.  Thus alternatively, one may try
to develop discrete state network, such as Hopfield network and
Boltzmann machine, model of a complex biological system using the
available information on the activation and repression from one
signaling molecules to another, e.g., the kind of signaling wiring
diagram like Fig. \ref{fig_diagram}.

Recently, Li, et.al. \cite{LLL04} have developed a discrete
deterministic Boolean model by applying the approach of Hopfield for
neural networks to the yeast cell-cycle regulatory network with 11
nodes according to Fig. \ref{fig_diagram}. They have found that the
topology of the network provides significant robustness of the
dynamics toward the check points, i.e., steady states.  The
deterministic Boolean model has been further extended to incorporate
stochastic dynamics \cite{ZQ1}. Several biologically interesting
results have been obtained; these include the numerical evidence for
the existence of a single dominant cell cycle and its robustness
under a large range of noise level.

   In the present paper, we will investigate the synchronized stochastic dynamics of
the discrete network model of yeast cell-cycle regulation where
stochasticity has been kept rather than being averaged out from the
start, applying the mathematical circulation theory of Markov
chains. By comparing the network dynamics of the stochastic model
with its corresponding deterministic network counterpart, we show
that the synchronized dynamics can be soundly characterized by a
dominant circulation in the stochastic model, which is the natural
generalization of the deterministic limit cycle in the deterministic
system.

    It is shown that a large separation in the magnitude of the circulations, between
a dominant, main cycle and the rest, gives rise to the stochastic
synchronization phenomenon, and the power spectrum of the trajectory
has a main peak, whose period converges perfectly to the number of
states in the dominant cycle.  Furthermore, the net circulation of
the dominant cycle increases monotonically with the noise-strength
parameter $\beta$, approaching to its deterministic limit. Together,
these observations provide a clear picture of the nature of the
synchronization in a stochastic network in terms of the
probabilistic circulation of NESS. The main part of the present
paper, Section IV presents our study on the synchronization and
dominant circulation in the cell-cycle network through numerical
computations.

    For the completeness of the work, we first give a theoretical sketch of some
relevant results on biological networks in Section II, including a
classification of the deterministic and stochastic Boolean networks
and their correspondence. Then a short review of the mathematical
theory of stochastic circulation for Markov chains is introduced and
applied to Hopfield network and Boltzmann machine in Section III. It
is shown that the stochastic Boolean network is reversible if and
only if the matrix $T$ in the model is symmetric, and the net NESS
circulation is strictly positive as long as the probability of the
directed cycle is larger than that of its reversal. Finally, the
conclusion and some further discussions are provided in Section V.

\section{Theoretical Sketch of Some Relevant Results on Biological
Networks: Hopfield networks, Boltzmann machines, and Their
Relations}

   We first give a brief account of the deterministic Hopfield networks
and its stochastic incarnation, the Boltzmann machines.  Since the
influential work of J.J. Hopfield in 1980s' \cite{H82,H84}, the
deterministic Boolean (Hopfield) network has been applied to various
fields of sciences. Amit \cite{Am} has introduced a temperature-like
parameter $\beta$ that characterizes the noise in the network and
constructed a probabilistic Boolean network called Boltzmann
machine. The Hopfield networks and Bolzmann machines have found wide
range of applications in biology. They can be mainly categorized
into several classes as follows \cite{Am}.

    Let us suppose $N$ is a fixed integer, $S=\{1,2,\cdots,N\}$. We
take the state space as $\{-1,1\}^S$.

\noindent {\bf Model A: deterministic}

{\it A1 (discrete time, synchronous, McCulloch-Pitts):} Denote the
state of the $n$th step as $X_n=(X_n^1,X_n^2,\cdots,X_n^N)$, then
the dynamic is
\begin{equation}
    X_{n+1}^i=\textrm{sign}\footnote{The function $\textrm{sign}(x)=\left\{\begin{array}{ll}1&x>0;\\0&x=0;\\-1&x<0.\end{array}\right.$}\left(
    \sum_{j=1}^N T_{ij}X_n^j-U_i\right), ~\textrm{if}~
    \sum_{j=1}^N T_{ij}X_n^j\neq U_i;
\end{equation}
and if $\sum_{j=1}^N T_{ij}X_n^j=U_i$, then $X_{n+1}^i$ randomly
choose $1$ or $-1$ with probability $\frac{1}{2}$ respectively.
$U_i~(1\le i\le N)$, given {\it a priori}, are called the threshold
of the $i$th unit.

{\it A2 (continuous time, synchronous):} Every state has an
exponentially distributed stochastic waiting-time, with mean
waiting-time $\lambda^{-1}$, then chooses the next state by the same
rule of model $A1$.

{\it A3 (discrete time, asynchronous, Hopfield):} The neurons are
updated one by one, in some prescribed sequence, or in a random
order. If the previous state $\sigma$ satisfies $\sum_{j=1}^N
T_{ij}\sigma_j>U_i$, then $\sigma_i$ changes to be $1$, otherwise
changes to be $-1$.

{\it A4 (continuous time, asynchronous):} Every state has an
exponentially distributed stochastic waiting-time, with rate
constant $\lambda$, then chooses the next state by the same rule of
model $A3$.

\begin{rem}
The $(-1,1)$ representation and the $(0,1)$ representation are
equivalent. Biologists usually use the $(0,1)$ representation in
modelling biological networks, where $0$ and $1$ denote the resting
and activated states, respectively.  Physicists usually use the
$(-1,1)$ representation for spin systems.
\end{rem}

\begin{rem}
Note that by deterministic, we mean the transition from one state to
the next is deterministic.  But the systems with continuous-time
will still behave stochastically due to the Poisson nature in the
transition time.
\end{rem}

\noindent {\bf Model B: stochastic Boltzmann machines}

{\it B1 (discrete time, synchronous):} Consider the Markov chain
\begin{equation}
    \{X_n=(X_n^1,X_n^2,\cdots,X_n^N), n = 0,1,2,\cdots\}
\end{equation}
on state space $\{-1,1\}^S$, with transition probability given as
follows: for each pair of states $\sigma,\eta\in \{-1,1\}^S$, the
probability transiting from $\sigma$ to $\eta$
\begin{equation}
  p_{\sigma\eta}=\prod_{i=1}^N \frac{\exp(\beta\eta_i(\sum_{j=1}^N T_{ij}\sigma_j-U_i)}
    {\exp(\beta(\sum_{j=1}^N T_{ij}\sigma_j-U_i))
    +\exp(-\beta(\sum_{j=1}^N T_{ij}\sigma_j-U_i))},
\end{equation}
where $\beta (>0)$, $T_{ij}$ and $U_i~ (1\le i,j\le N)$ are
parameters of the model.

{\it B2 (continuous time, synchronous):} Consider the
continuous-time Markov chain $\{\xi_t:t\geq 0\}$ on state space
$\{-1,1\}^S$.  Every state waits an exponential time with meantime
$\lambda^{-1}$ until choosing the next state by the rule of model
$B1$. So the transition density matrix is
\begin{equation}
    q_{\sigma\eta}=\lambda p_{\sigma\eta},
    ~\forall~\sigma,\eta\in \{-1,1\}^S.
\end{equation}

{\it B3 (discrete time, asynchronous):} Denote  $\sigma^i$ to be the
new state which changes the sign of the $i$th coordinate of
$\sigma$. Consider the Markov chain
$\{X_n=(X_n^1,X_n^2,\cdots,X_n^N), ~n=0,1,2,\cdots\}$ on state space
$\{-1,1\}^S$. The neurons are updated one by one, in some prescribed
sequence, or in a random order. Then choose the next state according
to the probability:
$$p_{\sigma\sigma^i}=\frac{\exp(-\beta\sigma_i(\sum_{j=1}^N T_{ij}\sigma_j-U_i))}{\exp(\beta(\sum_{j=1}^N T_{ij}\sigma_j-U_i))+\exp(-\beta(\sum_{j=1}^N T_{ij}\sigma_j-U_i))},~\sigma \in \{-1,1\}^S,$$
where $\beta>0$; and $p_{\sigma\eta}=0$, if $\eta\neq \sigma^i$ for
each $i$.

{\it B4 (continuous time, asynchronous):} Every state waits an
exponential time with meantime $\lambda^{-1}$ until choosing the
next state by the rule of model $B3$. So the transition density
matrix is
\begin{equation}
    q_{\sigma\sigma^i}=\lambda p_{\sigma\sigma^i},~\forall \sigma\in \{-1,1\}^S.
\end{equation}

    The third class given below is a variant of the
model $B1$.  It is included here since it is the model used for the
probabilistic Boolean network of cell-cycle regulation.

\noindent {\bf Model C: deformation}

Fix $\alpha>0$. Consider a new Markov chain
$\{X_n=(X_n^1,X_n^2,\cdots,X_n^N), n=0,1,2,\cdots\}$ on the state
space $\{-1,1\}^S$ taking the model $B1$ as defined initially, with
transition probability given as follows:
\begin{equation}
    P(X_{n+1}|X_n)=\prod_{i=1}^N P_i(X_{n+1}^i|X_n),
\end{equation}
where we define
\begin{eqnarray}
&&P_i(X_{n+1}^i|X_n)=\nonumber\\
&&\left\{\begin{array}{ll}\frac{\exp(\beta X_{n+1}^i(\sum_{j=1}^N
T_{ij}X_n^j-U_i))}
    {\exp(\beta(\sum_{j=1}^N T_{ij}X_n^j-U_i))+\exp(-\beta(\sum_{j=1}^N T_{ij}X_n^j-U_i))}
        &\textrm{if}~ \sum_{j=1}^N T_{ij}X_n^j\neq U_i\\
\frac{1}{1+e^{-\alpha}}
        &\textrm{if}~ \sum_{j=1}^N
T_{ij}X_n^j=U_i ~\textrm{and}~
X_{n+1}^i=X_n^i\\\frac{e^{-\alpha}}{1+e^{-\alpha}}
        &\textrm{if}~
\sum_{j=1}^N T_{ij}X_n^j=U_i ~\textrm{and}~ X_{n+1}^i=1-X_n^i.
\end{array}\right.\nonumber
\end{eqnarray}

\begin{rem}
The model C differs from B1 when $\sum_{j=1}^NT_{ij}X_n^j-U_i=0$,
and the latter is also a special case of the former when $\alpha=0$.
\end{rem}

\section{Mathematical Circulation Theory of Network Nonequilibrium Steady
States}

There are many different approaches to the theory of nonequilibrium
statistical mechanics in the past \cite{Ha,NP,Kei}, mathematical
theories of which have emerged in the last two decades, and Jiang
et.al. have summarized their results of this theory in a recent
monograph \cite{JQQ}. The most important concepts in the theory are
($i$) reversibility of a stationary process that corresponds to
thermodynamic equilibrium, and ($ii$) the circulation in a
stationary process which corresponds to NESS. A key result of the
theory is the circulation decomposition of NESS.

\subsection{Circulation theory of nonequilibrium steady states}

Hill \cite{hill} constructed a theoretical framework for discussions
of vivid metabolic systems, such as active transport, muscle
contractions, etc. The basic method of his framework is diagram
calculation for the cycle flux on the metabolic cycles of those
systems \cite{hill}. He successively found that the result from
diagram calculation agrees with the data of the numbers of
completing different cycles given by random test (Monte Carlo test),
but did not yet prove that the former is just the circulation rate
in the sense of trajectory of a corresponding Markov chain. In
Chapter 1 and 2 of \cite{JQQ}, Markov chains with discrete time and
continuous time parameter are used as models of Hill's theory on
circulation in biochemical systems. The circulation rate is defined
in the sense of trajectories and the expressions of circulation rate
are calculated which coincide with Hill's result obtained from
diagrams. Hence the authors verify that Hill's cycle flux is
equivalent to the circulation rate defined in the sense of
trajectories.

Below we only state the circulation theory of discrete-time Markov
chains, and refer to \cite[Chapter 2]{JQQ} for the quite similar
circulation theory of continuous-time Markov chains.

 First of all, we state the main results, rewritten in terms
of the cycle representation of stationary homogeneous Markov chains
(\cite{SK}, Theorem 1.3.1), which is analogous to the Kirchoff's
current law and circuit theory in the networks of master equation
systems \cite{Sc}.

\begin{thm}
Given a finite oriented graph $G=(V,E)$ and the weight on every edge
$\{\omega_e>0: e\in E\}$. If the weight satisfies the balance
equation
\begin{equation}\label{balance-eq}
\sum_{e^+=i} \omega_e=\sum_{e^-=i} \omega_e,~\forall i\in V,
\end{equation}
then there exists a positive function defined on the oriented cycles
$\{\omega_c: c=(e_1,e_2,\cdots,e_k), k\in Z^+\}$, such that
\begin{equation}
\omega_e=\sum_{e\in c} \omega_c,~\forall e\in E.
\end{equation}
\end{thm}

For a finite stationary Markov chain $X$ with finite state space
$S=\{1,2,\cdots,N\}$, transition matrix $P=\{p_{ij}:~ i,j\in S\}$
and invariant distribution $\bar{\pi}=\{\pi_1,\pi_2,\cdots,\pi_N\}$,
one can take its state space to be the group of vertexes, and
$E=\{e: e^+=i,e^-=j,p_{ij}>0\}$ with weight
$\{\omega_e=\pi_ip_{ij}\}$. Then notice that (\ref{balance-eq}) is
satisfied, since $\sum_{j}p_{ij}=1$  for each $i$ and
$$\sum_{j}\pi_ip_{ij}=\pi_i=\sum_j \pi_jp_{ji},~\forall i\in V,$$
we can conclude that

\begin{cor} (cycle decomposition)
\label{cycdecomp} For an arbitrary finite stationary Markov chain,
there exists a positive function defined on the group of oriented
circuits $\{\omega_c: c=(i_1,i_2,\cdots,i_k), k\in Z^+\}$ such that
\begin{equation}
\pi_ip_{ij}=\sum_{c} \omega_cJ_c(i,j),~\forall i,j\in V,
\end{equation}
where $J_c(i,j)$ is defined to be 1 if the cycle c includes the path
$i\rightarrow j$, otherwise 0.
\end{cor}

\begin{defn}
$\omega_c$ is called the {\bf circulation} along cycle c.
\end{defn}

For any $i,j\in S, i\not=j$,
\begin{equation}
 \pi_i p_{ij}-\pi_j p_{ji}
  =\sum_{c} (w_c-w_{c_-}) J_c(i,j),
 \label{circ-decomp-con}
\end{equation}
where $c_-$ denotes the reversed cycle of $c$. Equation
(\ref{circ-decomp-con}) is called the {\bf circulation
decomposition} of the stationary Markov chain $X$.

It can be proved that generally the circulation decomposition is not
unique, i.e. it is possible to find another set of cycles ${\cal C}$
and weights on these cycles $\{w_c|c\in {\cal C} \}$ which also fit
(\ref{circ-decomp-con}).

However, the most reasonable choice of circulation definition is the
one defined in the sense of trajectories form the probabilistic
point of view. Along almost every sample path, the Markov chain
generates an infinite sequence of cycles, and if we discard every
cycle when it is completed and at the meantime record it down, then
we can count the number of times that a specific cycle $c$ is formed
by time $t$, which we denote by $w_{c,t}(\omega)$.

The following theorem is recapitulated from \cite[Theorem
1.3.3]{JQQ}.
\begin{thm}\label{Thm-circulation}
Let ${\cal C}_n(\omega)$, $n=0,1,2,\cdots$, be the class of all
cycles occurring until $n$ along the sample path $\{X_l(\omega)\}$.
Then the sequence $({\cal C}_n(\omega),w_{c,n}(\omega)/n)$ of sample
weighted cycles associated with the chain $X$ converges almost
surely to a class $({\cal C}_{\infty},w_c)$, that is,
\begin{equation}
 {\cal C}_{\infty}=\lim_{n\rightarrow +\infty} {\cal C}_n(\omega), {\rm ~~a.e.}
\end{equation}
\begin{equation}
 w_c=\lim_{n\rightarrow +\infty} \frac{w_{c,n}(\omega)}{n}, {\rm ~~a.e.}
\end{equation}
Furthermore, for any directed cycle
 $c=(i_1,i_2,\cdots,i_s)\in{\cal C}_{\infty}$, the weight $w_c$ is given by
\begin{equation}
 w_c=p_{i_1i_2}p_{i_2i_3}\cdots p_{i_{s-1}i_s}p_{i_si_1}
      \frac{D(\{i_1,i_2,\cdots,i_s\}^c)}
           {\sum_{j\in S} D(\{j\}^c)}.
\end{equation}
where $D=\{d_{ij}\}=I-P=\{\delta_{ij}-p_{ij}\}$ and $D(H)$ denotes
the determinant of D with rows and columns indexed in the index set
H. The function $\delta_{ij}=\left\{\begin{array}{ll}0,&i\neq
j;\\1,&i=j,\end{array}\right.$ is the well known Kronecker delta
function.
\end{thm}

It is important to emphasize that the circulations defined in the
above theorem also satisfy the circulation decomposition relation
(\ref{circ-decomp-con}).

The above theorem not only rigorously confirms the Hill's theory,
but also gives a prior substitute method of the widely used
diagrammatic method. The complexity of directed diagrams and cycles
increases rapidly with the number of states in the model, while the
determinant interpretation is much more systematic and easy to be
applied using the mathematics softwares. But it will still cost
excessive time to compute these determinants if there are hundreds
of states in the model.

Luckily, a Monte Carlo method using the so-called {\bf derived chain
method} \cite[Section 1.2]{JQQ} to compute the cycle fluxes has
already been developed according to the above theorem, the main idea
of which is just simply to discard every cycle when it is completed
and at the meantime record it down so as to count the number of
times that a specific cycle $c$ is formed by time $t$(i.e.
$w_{c,t}(\omega)$). Then when the time $t$ is long enough, one gets
the approximated circulation of the cycle $c$ (i.e.
$w_c\approx\frac{w_{c,t}(\omega)}{t}$).

Actually, the Boolean yeast cell-cycle network model discussed in
the next section has $2048$ states and we find that the Monte Carlo
method is much more efficient than the method of determinant
interpretation, because the latter is even impossible to be applied
to such a large model upon a normal computer.

The relationship between circulation and NESS is as follows, which
is recapitulated from \cite[Theorem 1.4.8]{JQQ}.
\begin{thm}\label{rev-cir-Kol}
Suppose that $X$ is an irreducible and positive-recurrent stationary
Markov chain with the countable state space $S$, the transition
matrix $P=(p_{ij})_{i,j\in S}$ and the invariant probability
distribution $\Pi=(\pi_i)_{i\in S}$, and let $\{w_c:c\in{\cal
C}_{\infty}\}$ be the circulation distribution of $X$, then the
following statements are
equivalent: \\
(i) The Markov chain $X$ is reversible. \\
(ii) The Markov chain $X$ is in detailed balance, that is,
 $$\pi_i p_{ij}=\pi_j p_{ji},\forall i,j\in S.$$
(iii) The transition probability of $X$ satisfies the {\bf
Kolmogorov cyclic condition}:
 $$p_{i_1i_2}p_{i_2i_3}\cdots p_{i_{s-1}i_s}p_{i_si_1}
   =p_{i_1i_s}p_{i_si_{s-1}}\cdots p_{i_3i_2}p_{i_2i_1},$$
for any directed cycle $c=(i_1,\cdots,i_s)$. \\
(iv) The components of the circulation distribution of $X$ satisfy
the symmetry condition:
 $$w_c=w_{c_-},\forall c\in{\cal C}_{\infty}.$$
\end{thm}

Consequently, when the system is in a nonequilibrium steady state,
there exists at least one cycle, containing at least three states,
round which the circulation rates of one direction and its opposite
direction are asymmetric (unequal), so as to cause a net circulation
on the cycle. In theoretic analysis, if there is  a large separation
in the magnitude of the circulation, between few dominant, main
cycles and the rest, it gives rise to the stochastic synchronization
phenomenon and helps to distinguish the most important main
biological pathways, which can be observed in experiments.

\subsection{Applied to Hopfield network and Boltzmann machine}

The keys to understand synchronization behavior in stochastic models
are ($i$) establishing a correspondence between a stochastic
dynamics and its deterministic counterpart; and ($ii$) identifying
the cyclic motion in the stochastic models.

In the framework of the stochastic theory, deterministic models are
simply the limits of stochastic processes with vanishing noise. This
is best illustrated in the following proposition.

\begin{prop}
\label{prop_stod} With the same initial distribution, when
$\beta\rightarrow\infty$, the model $B_k$ converges to the model
$A_k$ in distribution, for $k=1,2,3,4$.
\end{prop}
{\bf Proof:} From
\[
\lim_{\beta\rightarrow\infty}\frac{\exp(\beta\eta_i(\sum_{j=1}^N
T_{ij}\sigma_j-U_i))}{\exp(\beta(\sum_{j=1}^N
T_{ij}\sigma_j-U_i))+\exp(-\beta(\sum_{j=1}^N T_{ij}\sigma_j-U_i))}
\]
\begin{equation}
=\left\{\begin{array}{ll} 1,&\eta_i(\sum_{j=1}^N
T_{ij}\sigma_j-U_i)>0;\\0,&\eta_i(\sum_{j=1}^N
T_{ij}\sigma_j-U_i)<0;\\\frac{1}{2},&\sum_{j=1}^N
T_{ij}\sigma_j=U_i.\end{array}\right.\qed
\end{equation}

  We shall further discuss the necessary condition for stochastic
Boolean networks to have cyclic motion. In the theory of neural
networks, one of Hopfield's key results \cite{H82,H84} is that for
symmetric matrix $T_{ij}$, the network has an energy function.  In
the theory of Markov processes, having a potential
function(Kolmogorov cyclic condition in Theorem \ref{rev-cir-Kol})
is a sufficient and necessary condition for a reversible process. A
connection is established in the following theorem.

\begin{thm}\label{thm-ness}
For $k=1,2,3,4$, the Markov chain $\{X_n\}$ in the model $B_k$ is
reversible if and only if $T_{ij}=T_{ji}, \forall i,
j=1,2,\cdots,N$, i.e. the matrix $T$ is symmetric.
\end{thm}
{\bf Proof:} According to theorem \ref{rev-cir-Kol}, we can use the
Kolmogorov cyclic condition to complete the proof. Since every two
states are communicative, one only needs to consider the cycles
consist of three different states.

Denote these states by $\alpha=(\alpha_1,\alpha_2,\cdots,\alpha_N)$,
$\sigma=(\sigma_1,\sigma_2,\cdots,\sigma_N)$, and
$\eta=(\eta_1,\eta_2,\cdots,\eta_N)$. Consider the cycle
$\alpha\rightarrow\sigma\rightarrow\eta\rightarrow\alpha$ then

\begin{eqnarray}\label{eq1}
&&\frac{p_{\alpha\sigma}p_{\sigma\eta}p_{\eta\alpha}}{p_{\alpha\eta}p_{\eta\sigma}p_{\sigma\alpha}}\nonumber\\
&=&\left[\prod_{i=1}^N \frac{\exp(\beta\sigma_i(\sum_{j=1}^N T_{ij}\alpha_j-U_i))}{\exp(\beta(\sum_{j=1}^N T_{ij}\alpha_j-U_i))+\exp(-\beta(\sum_{j=1}^N T_{ij}\alpha_j-U_i))} \right.\nonumber\\
&&\times
\prod_{i=1}^N \frac{\exp(\beta\eta_i(\sum_{j=1}^N T_{ij}\sigma_j-U_i))}{\exp(\beta(\sum_{j=1}^N T_{ij}\sigma_j-U_i))+\exp(-\beta(\sum_{j=1}^N T_{ij}\sigma_j-U_i))}\nonumber\\
&&\left.\times \prod_{i=1}^N \frac{\exp(\beta\alpha_i(\sum_{j=1}^N T_{ij}\eta_j-U_i))}{\exp(\beta(\sum_{j=1}^N T_{ij}\eta_j-U_i))+\exp(-\beta(\sum_{j=1}^N T_{ij}\eta_j-U_i))}\right]\nonumber\\
&&\div\left[\prod_{i=1}^N \frac{\exp(\beta\eta_i(\sum_{j=1}^N T_{ij}\alpha_j-U_i))}{\exp(\beta(\sum_{j=1}^N T_{ij}\alpha_j-U_i))+\exp(-\beta(\sum_{j=1}^N T_{ij}\alpha_j-U_i))}\right.\nonumber\\
&&\times \prod_{i=1}^N \frac{\exp(\beta\sigma_i(\sum_{j=1}^N T_{ij}\eta_j-U_i))}{\exp(\beta(\sum_{j=1}^N T_{ij}\eta_j-U_i))+\exp(-\beta(\sum_{j=1}^N T_{ij}\eta_j-U_i))}\nonumber\\
&&\left.\times \prod_{i=1}^N \frac{\exp(\beta\alpha_i(\sum_{j=1}^N T_{ij}\sigma_j-U_i))}{\exp(\beta(\sum_{j=1}^N T_{ij}\sigma_j-U_i))+\exp(-\beta(\sum_{j=1}^N T_{ij}\sigma_j-U_i))}\right]\nonumber\\
&=&\exp\left\{-\beta\sum_{i,j} T_{ij}[(\alpha_i\sigma_j+\sigma_i\eta_j+\eta_i\alpha_j)-(\alpha_j\sigma_i+\sigma_j\eta_i+\eta_j\alpha_i)]\right\}\nonumber\\
&=&\exp\left[-\beta\sum_{i,j}
(T_{ij}-T_{ji})(\alpha_i\sigma_j+\sigma_i\eta_j+\eta_i\alpha_j)\right].
\end{eqnarray}

\noindent necessity: Let $\eta_k=-1, \forall k=1,2,\cdots,N$,
$\alpha_k=-1, \forall k\neq i$, $\alpha_i=1$, and $\sigma_k=-1,
\forall k\neq j$, $\sigma_j=1$. So $\sum_{k,l}
(T_{kl}-T_{lk})(\alpha_k\sigma_l+\sigma_k\eta_l+\eta_k\alpha_l)=2(T_{ij}-T_{ji}),$
if the Markov chain is reversible then by Theorem
\ref{rev-cir-Kol}(iii)
$p_{\alpha\sigma}p_{\sigma\eta}p_{\eta\alpha}=p_{\alpha\eta}p_{\eta\sigma}p_{\sigma\alpha}$,
we can conclude that $T_{ij}=T_{ji}$, $\forall i,j=1,2,\cdots,N$.

\noindent sufficiency: If $T_{ij}=T_{ji}, \forall i,
j=1,2,\cdots.N$, then from (\ref{eq1}) one has
$p_{\alpha\sigma}p_{\sigma\eta}p_{\eta\alpha}
    =p_{\alpha\eta}p_{\eta\sigma}p_{\sigma\alpha},~\forall
\alpha,\sigma,\eta$.\qed

From the above two conclusions, it is obvious that
\begin{cor}\label{Cor1}
When $T_{ij}=T_{ji}, \forall i, j=1,2,\cdots,N$, there doesn't exist
any limit cycle consist of more than two states in model $A_k$,
$k=1,2,3,4$.
\end{cor}

\begin{rem}
Part of the proof for the corollary \ref{Cor1} can be found in the
literature, such as \cite{YZ}. The proof we give here, however,
provides a new point of view from that of stochastic convergence,
and is complete.
\end{rem}

\begin{lem}\label{lem}
For $k=1,2,3,4$, the Markov chain $\{X_n\}$ in the model $B_k$ is
reversible when $\beta$ is zero, and the Markov chain $\{X_n\}$ in
the model $C$ is reversible when $\beta$ and $\alpha$ are both zero.
\end{lem}

\emph{Proof:} In the case when $\beta$ and $\alpha$ are both zero,
then for arbitrary two states $\sigma$ and $\eta$, we have
$p_{\sigma\eta}=\frac{1}{2}$. So the Markov chain $\{X_n\}$ in the
model $B_k$, $k=1,2,3,4$, and model $C$ satisfies the Kolmogorov
cyclic condition. \qed

\section{Synchronized Dynamics in a Stochastic Network of Yeast Cell-Cycle Regulation}

    From biochemical perspective, the microscopic variables
for a cellular regulatory network are the concentrations, or
numbers, of various mRNAs, regulatory proteins, and cofactors. If
all the biochemistry were known, then the dynamics of such a network
would be represented by a chemical master equation \cite{Gi,Mc}.
Unfortunately, much of the required information is not available,
nor such a ``fully-detailed'' model will always be useful.
Phenomenologically the concentrations of key players of a
biochemical regulatory network can often be reduced to two or three
states, such as resting state, activated state, inactivated state,
etc. \cite{FMWT}.  The interactions between these states are usually
determined from experimental data.

\subsection{The stochastic network model of Zhang et. al.}

    We now turn to the stochastic model of cell-cycle
network as developed by Li et.al. \cite{LLL04} and Zhang et.al.
\cite{ZQ1,ZQ2}.  The model in \cite{LLL04} is a deformation of model
$A_1$ with the $(0,1)$ representation instead of the $(-1,1)$
representation: Denote the state of $n$-th step as
$X_n=(X_n^1,X_n^2,\cdots,X_n^N)$, then the dynamic is
\begin{equation}
    X_{n+1}^i=\frac{1}{2}\left[\textrm{sign}\left(
        \sum_{j=1}^N T_{ij}X_n^j\right)+1\right],
    ~\textrm{if}~\sum_{j=1}^N T_{ij}X_n^j\neq 0;
\end{equation}
while $X_{n+1}^i=X_{n}^i$ if $\sum_{j=1}^N T_{ij}X_n^j=0$.

The main results of \cite{LLL04} are that the network is both
dynamically and structurally stable.  The biological steady state,
known as the G1 phase of a cell cycle, is a global attractor of the
dynamics; the biological pathway, i.e., the returning to G1 phase
after perturbation, is in a globally attracting basin. There is no
limit cycle in this model.

Zhang et.al. \cite{ZQ1} implemented a probabilistic Boolean network
for the cell-cycle protein interaction network.  They found that
both the biological steady state and the biological pathway are well
preserved under a wide range of noise level. The model in \cite{ZQ1}
is model $C$, with the trivial difference of taking the $(0,1)$
representation rather than $(-1,1)$ representation.


Notice that, according to Proposition \ref{prop_stod}, when
$\beta,\alpha\rightarrow\infty$, this model recovers the
deterministic model in \cite{LLL04}; hence, it is implicit that
there is no dominant cycle with significant circulation when $\beta$
and $\alpha$ are large.

There is an important difference between the previous models and the
most recently developed one in \cite{ZQ2}, which will be
investigated currently in more detail. The earlier works all treat
the cell mass (volume), implicitly through protein Cln3 (node 1), as
a parameter of their models rather than a dynamic part of the
models. Under this setting, the G1 phase of the cell cycle,
$(0,0,0,0,1,0,0,0,1,0,0)$ in our binary representation, is a global
attractor. The biological cell cycle, however, has a clear sense of
cycling with direction. It is a time-irreversible process
accompanied with positive entropy production; it is a system in
NESS.  The cell mass has been incorporated into the transition
dynamics in \cite{ZQ2}, again implicitly characterized by the
$\gamma$ parameter: When the system is in G1 phase, a signal to exit
the G1 phase by activating Cln3 (node 1), known as START in cell
biology, will come with a probability given in Eq. (\ref{eq4gamma})
below.  It represents the trigger implicitly associated with cell
mass. This is the idea of \cite{ZQ2}, but the method of attack used
there is ad hoc with no complete picture.

In the present paper, we focus on the model in \cite{ZQ2}, thus, is
a deformation of model $C$: fix $\alpha>0,~\gamma>0$, consider the
Markov chain $\{X_n=(X_n^1,X_n^2,\cdots,X_n^N):n=0,1,2,\cdots\}$ on
the state space $\{0,1\}^S$, whose transition probability is as
follows:
$$P(X_{n+1}|X_n)=\prod_{i=1}^N P_i(X_{n+1}^i|X_n),$$
where
$$P_i(X_{n+1}^i|X_n)=\frac{\exp(\beta (2X_{n+1}^i-1)(\sum_{j=1}^N T_{ij}X_n^j))}{\exp(\beta(\sum_{j=1}^N T_{ij}X_n^j))+\exp(-\beta(\sum_{j=1}^N T_{ij}X_n^j))},$$
if $\sum_{j=1}^N T_{ij}X_n^j\neq 0$.

$$P_i(X_{n+1}^i|X_n)=\left\{\begin{array}{ll}\frac{1}{1+e^{-\alpha}},&~X_{n+1}^i=X_n^i,\\\frac{e^{-\alpha}}{1+e^{-\alpha}},&~X_{n+1}^i=1-X_n^i,
\end{array}\right.$$
if $\sum_{j=1}^N T_{ij}X_n^j=0$ and
$X_n\neq(0,0,0,0,1,0,0,0,1,0,0)$ or $i\geq 2$.

\begin{equation}
 P_1(X_{n+1}|X_n)=\left\{\begin{array}{ll}\frac{1}{1+e^\gamma},&~X_{n+1}^1=X_n^1,\\\frac{e^\gamma}{1+e^\gamma},&~X_{n+1}^1=1-X_n^1,
\end{array}\right.
\label{eq4gamma}
\end{equation}
if $\sum_{j=1}^N T_{1j}X_n^j=0$ and $X_n=(0,0,0,0,1,0,0,0,1,0,0)$.

The work \cite{ZQ2} only introduced this model which is more
reasonable than that in \cite{ZQ1}, but the results and conclusions
in \cite{ZQ2} are very short and shallow. So the aim of the present
paper is to give more profound and comprehensive insight into their
work, applying the mathematical theory of nonequilibrium steady
states.

The matrix $T$ in all the models above \cite{LLL04,ZQ1,ZQ2}
according to Fig.\ref{fig_diagram} is
\begin{equation}\label{T-matrix}
T=\left [\begin{array}{lllllllllll}
-0.1&0&0&0&0&0&0&0&0&0&0\\
1&0&0&0&0&0&0&0&0&-1&0\\
1&0&0&0&0&0&0&0&0&-1&0\\
0&0&1&-0.1&0&0&0&0&0&0&0\\
0&0&0&-1&0&0&1&-1&0&-1&0\\
0&0&0&0&0&-0.1&1&0&0&-1&1\\
0&0&0&0&-0.1&0&0&0&0&1&1\\
0&1&0&0&0&0&-1&0&-1&0&0\\
0&0&0&-1&0&1&1&-1&0&-1&0\\
0&0&0&0&-0.1&0&-1&1&-1&0&1\\
0&0&0&0&0&0&0&1&0&1&-0.1
\end{array}\right],
\end{equation}
where $T_{ij}=1$ for a positive regulation of protein $i$ to protein
$j$ and $T_{ij}=-1$ for a negative regulation of protein $i$ to
protein $j$. If the protein $i$ has a self-degradation loop,
$T_{ii}=-0.1$.

This matrix is not symmetric. Hence, according to theorem
\ref{thm-ness}, the stochastic dynamics has a NESS with nonzero
circulation.

Notice that, when $\beta$, $\alpha$ and $\gamma\rightarrow\infty$,
this model converges to a deterministic model similar to that in
\cite{LLL04}: Denote the state of n-th step as
$X_n=(X_n^1,X_n^2,\cdots,X_n^N)$, then the dynamic is
\begin{equation}
    X_{n+1}^i=\frac{1}{2}
    \left[\textrm{sign}\left(
    \sum_{j=1}^N T_{ij}X_n^j\right)+1\right],
    ~\textrm{if}~\sum_{j=1}^N T_{ij}X_n^j\neq 0;
\end{equation}
while $X_{n+1}^i=X_{n}^i$ if $\sum_{j=1}^N T_{ij}X_n^j=0$ and
$X_n\neq(0,0,0,0,1,0,0,0,1,0,0)$ or $i\geq 2$;
$X_{n+1}^1=1-X_{n}^1$ if $\sum_{j=1}^N T_{1j}X_n^j=0$ and
$X_n=(0,0,0,0,1,0,0,0,1,0,0)$.

The corresponding deterministic model has a limit cycle consist of
13 state.  So there is also a main cycle whose circulation is
dominant when $\beta$, $\alpha$ and $\gamma$ are large, which is
described by Tab. \ref{tab1}.

\begin{table}[h]
\tiny{
\begin{center}
\begin{tabular}{ccccccccccccc}
Cln3&MBF&SBF&Cln1,2&Cdh1&Swi5&Cdc20/Cdc14&Clb5,6&Sic1&Clb1,2&Mcm1/SFF&Phase\\
\hline
(1&0&0&0&1&0&0&0&1&0&0)&Start\\
(0&1&1&0&1&0&0&0&1&0&0)&$G_1$\\
(0&1&1&1&1&0&0&0&1&0&0)&$G_1$\\
(0&1&1&1&0&0&0&0&0&0&0)&$G_1$\\
(0&1&1&1&0&0&0&1&0&0&0)&$S$\\
(0&1&1&1&0&0&0&1&0&1&1)&$G_2$\\
(0&0&0&1&0&0&1&1&0&1&1)&$M$\\
(0&0&0&0&0&1&1&0&0&1&1)&$M$\\
(0&0&0&0&0&1&1&0&1&1&1)&$M$\\
(0&0&0&0&0&1&1&0&1&0&1)&$M$\\
(0&0&0&0&1&1&1&0&1&0&0)&$M$\\
(0&0&0&0&1&1&0&0&1&0&0)&$G_1$\\
(0&0&0&0&1&0&0&0&1&0&0)&$G_1$
\end{tabular}
\end{center}
}
\caption[tab1]{The synchronous sequence of 13 states as recorded
in Li et al. \cite{LLL04}} \label{tab1}
\end{table}

\subsection{Synchronization and dominant circulation in
cell-cycle network}

While Zhang et.al. \cite{ZQ1,ZQ2} have introduced the probabilistic
Boolean network model of yeast cell cycle and tried to describe its
nonequilibrium dynamics, they are not fully aware of the important
interplay between the the nonequilibrium circulations of this model
and the observed synchronization phenomenon.

Furthermore, from numerical computation of the maximal probability
flux on cycles, the previous work \cite{ZQ1,ZQ2} seems to suggest
that the cycle with positive circulation is unique.  This is not the
case. Equipped with Corollary \ref{cycdecomp} on cycle decomposition
, we shall show that there are many other cycles with positive
circulation in the model, although their probability weights are
indeed quite small in contrast to the main cycle.  This large
separation in the magnitudes of the weights gives rise to the
stochastic synchronization. Therefore, this is indeed a
characterization of the stochastic synchronized dynamics.

Indeed, from the theory of nonequilibrium circulation of
discrete-time homogeneous finite Markov chains introduced in Section
III, the net circulation is strictly positive as long as the
probability of directed cycle is larger than that of its reversal.
This is true for any cycle on which the Kolmogorov cyclic condition
breaks down \cite[Theorem 1.4.8]{JQQ}, e.g., in the model in
\cite{ZQ2}, let states $\mu=(0,0,1,0,1,0,0,0,1,0,0)$,
$\sigma=(1,0,0,1,0,0,0,0,0,1,0)$, and $\nu=(0,1,0,0,0,0,1,0,0,0,1)$.
Consider the cycle
$\mu\rightarrow\sigma\rightarrow\nu\rightarrow\mu$. Then
$$\frac{p_{\mu\sigma}p_{\sigma\nu}p_{\nu\mu}}{p_{\mu\nu}p_{\nu\sigma}p_{\sigma\mu}}=e^{3\alpha+10.9\beta}\neq 1.$$
Then the net circulation of this cycle is not zero.

Numerical computations with the famous Gillespie's method \cite{Gi}
are carried out, and the results are given in the following figures.
The network with 11 binary nodes has a total of 2048 number of
states. Following Amit (\cite{Am}, pp. 75-79), we present the
dynamics of the network in terms of integers $0,1,2,\cdots,2047$,
where the state $(s_1,s_2,\cdots,s_{11})$ corresponds one-to-one to
$\sum_{i=1}^{11}s_i2^{i-1}$. This 1-d system obtained is reversible
if and only if the 11-d system is reversible.

\begin{figure}[h]
\centerline{\includegraphics[width=5in,height=4in]{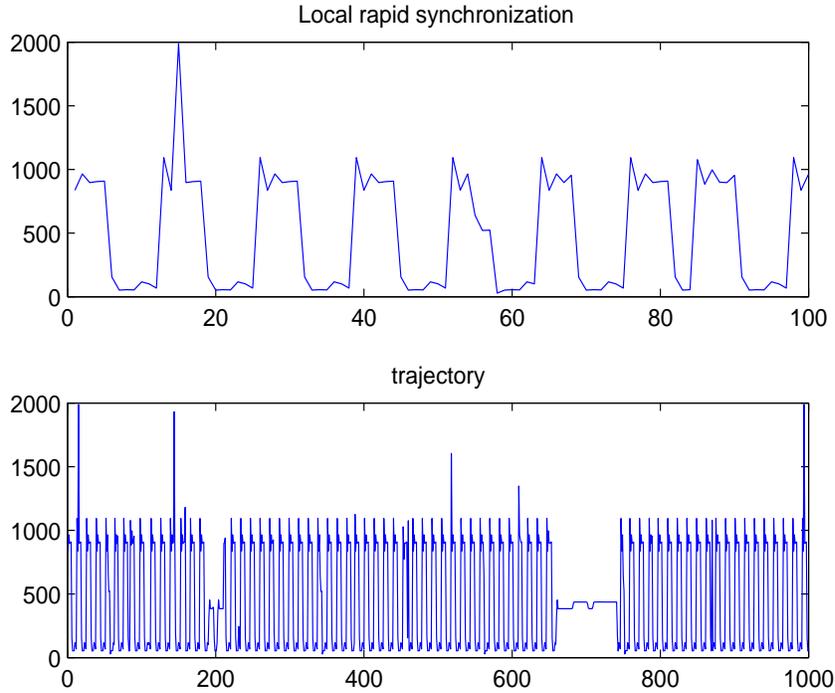}}
\caption[fig31]{Stochastic trajectory and synchronization.
Simulations are carried out with the parameters $\alpha=5$,
$\beta=6$, and $\gamma=5$.}
\label{fig31}
\end{figure}

Fig. \ref{fig31} is the basic behavior of a random trajectory.
The upper panel shows that there arises the phenomenon of local
rapid synchronization like that observed in \cite{H95}
during a very short time period, when $\beta$ is sufficiently
large.  The lower panel is a random trajectory over a much longer
time.

\begin{figure}[h]
\centerline{\includegraphics[width=4.5in,height=3in]{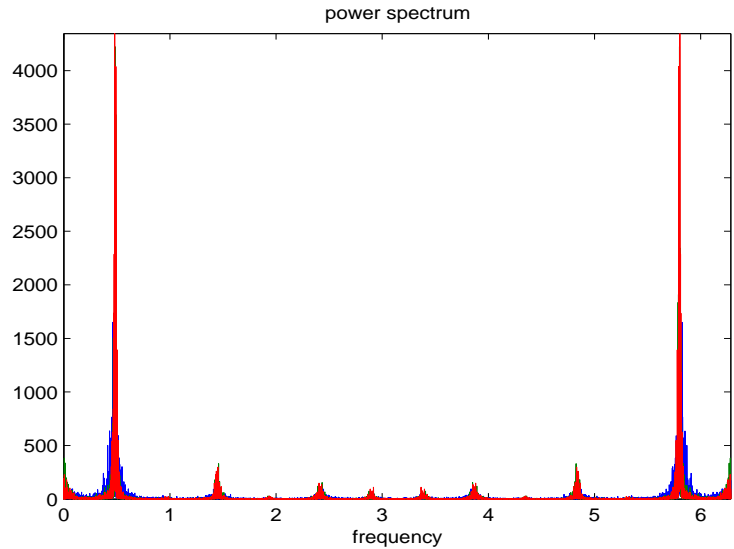}}
\caption[fig32]{Power spectrum of the overall trajectory with the
parameters $\alpha=5$, $\gamma=5$ fixed, and different $\beta$:
blue: $\beta$=2.4; green: $\beta$=4.8; red: $\beta=6$.  The discrete
Fourier transform causes an alias; hence the spectrum is
symmetric with respect to $\pi$ on the $[0,2\pi]$ interval.
}
\label{fig32}
\end{figure}

    To further characterize the synchronized dynamics,
Fig. \ref{fig32} shows the Fourier power spectrum of the
trajectories in Fig. \ref{fig31}.  Using MATLAB, the discrete
Fourier transform for time series  $\{x_1,x_2,\cdots,x_n\}$ is
defined as
\begin{equation}
    y_m=\left|\sum_{k=1}^n x_ke^{-i(2\pi/n)(m-1)(k-1)}\right|,
\end{equation}
which is just the magnitude of frequency $\frac{m-1}{n}2\pi$, $1\le
m\le n$.

Therefore, by the Herglotz theorem (\cite{QG}, p. 331), the power
spectrum of discrete trajectories has a symmetry $y_m=y_{n+2-m}$.
For different sets of parameters, all the calculations give the
distinguishable main peak in the Fig. \ref{fig32}. It is important
to mention that different trajectories tend to the same Fig.
\ref{fig32} by ergodicity.

\begin{figure}[h]
\centerline{\includegraphics[width=5in,height=4in]{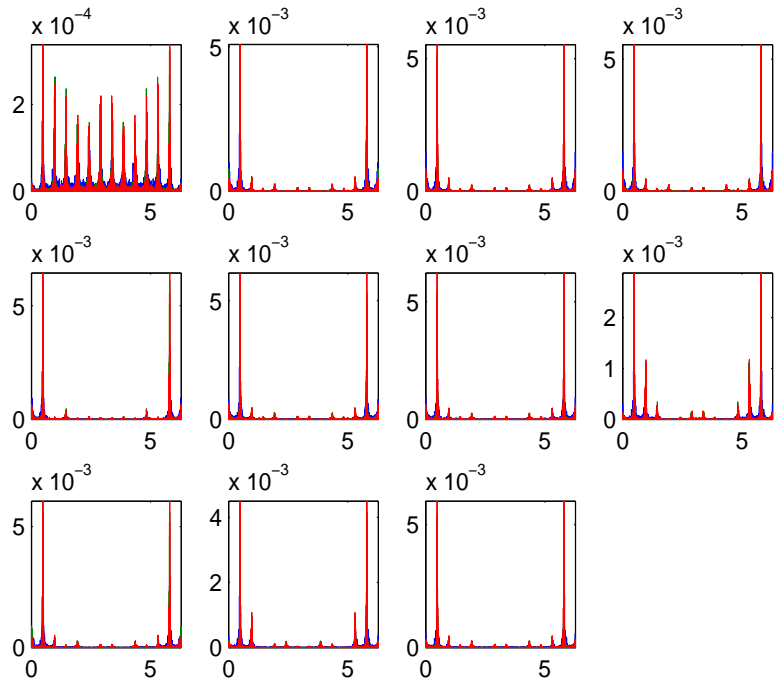}}
\caption[fig33]{Power spectra of individual nodes show a
synchronization among all the nodes with the parameters
$\alpha=5$, $\gamma=5$ fixed, and different $\beta$: blue:
$\beta$=2.4; green: $\beta$=4.8; red: $\beta$=6.}
\label{fig33}
\end{figure}

  The single dominant peak implies there exists a global
synchronization phenomenon.   Besides the main peak, the power
spectra also have several other smaller peaks, which correspond to
several cyclic motions with lower magnitudes.  Note that the
synchronized behavior is preserved in representation that maps,
one-to-one, from the 11 binary nodes to the integers $0-2047$.  It
is possible that the map will cause some distortion in the power
spectrum.  To further illustrate the synchronized behavior, Fig.
\ref{fig33} shows the power spectra of all the 11 individual nodes
in the network.  While subtle details are different, all exhibit the
dominant peak, similar to that of the overall dynamics. This
demonstrates further the synchronized dynamics in the network.

\begin{figure}[h]
\centerline{\includegraphics[width=6in,height=4in]{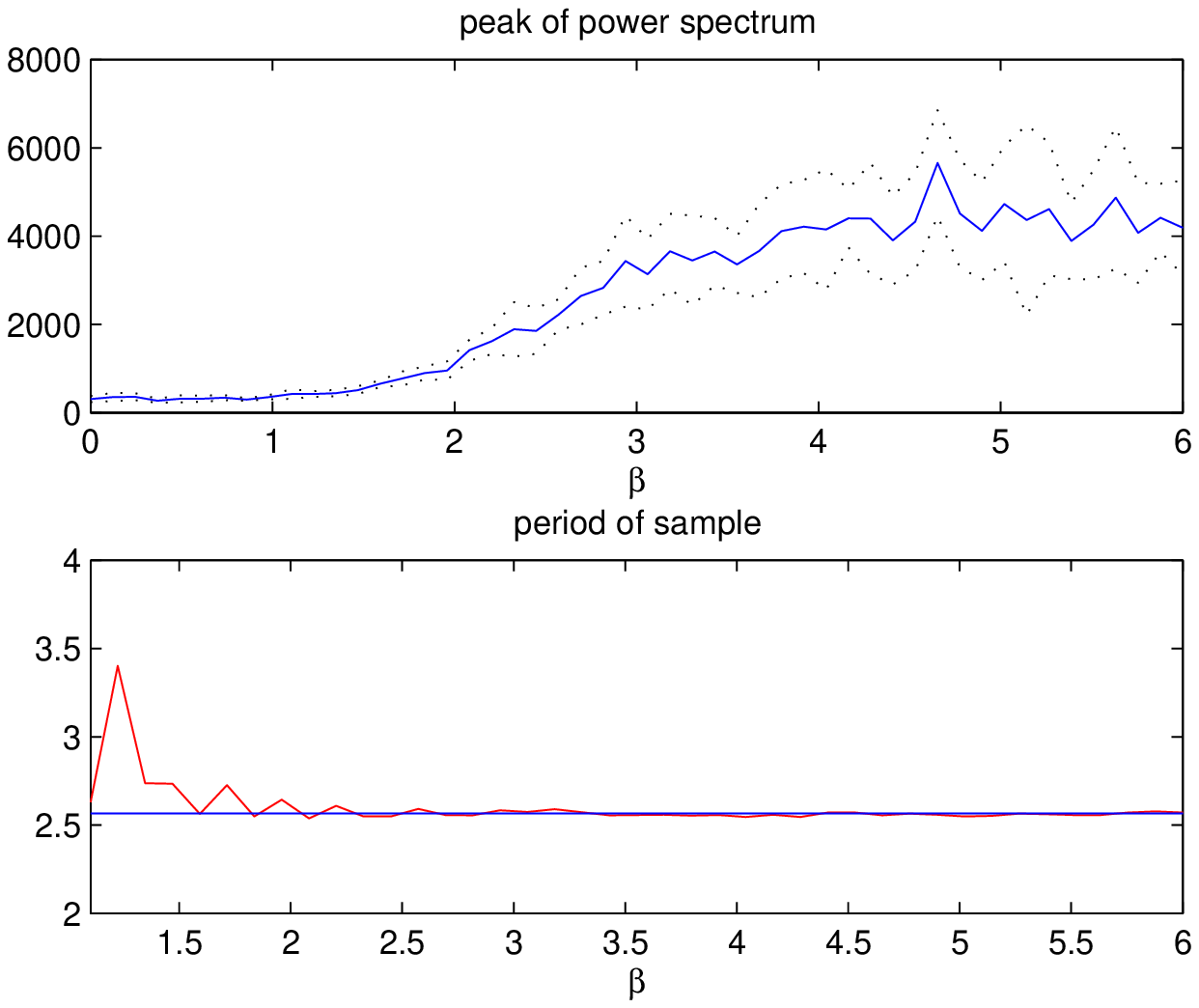}}
\caption[fig34]{Magnitude and period of the dominant
power-spectral peak as functions of $\beta$, with $\alpha=5$ and
$\gamma=5$. In the upper panel, the magnitude of the
dominant power-spectral peak is averaged over 10 simulations.
The solid curve is the mean, and the dotted curves are the mean
$\pm$ standard deviation.
In the lower panel, the abscissa for the period of the cycle,
$T$, is in logarithmic scale.  The period approaches to 13
(the horizonal line) with increasing $\beta$.}
\label{fig34}
\end{figure}

Fig. \ref{fig34} plots the magnitude and the period of the
dominant peak of the power spectrum in Fig. \ref{fig32}, as
functions of the noise strength, i.e., the parameter $\beta$. It
shows, as we have predicted, that the period converges to 13 which
corresponds perfectly to the number of states in the main cycle
when $\beta$ is large. We also put error bars on the upper panel
of Fig. \ref{fig34} with various values of $\beta$.

\begin{figure}[h]
\centerline{\includegraphics[width=4in,height=3in]{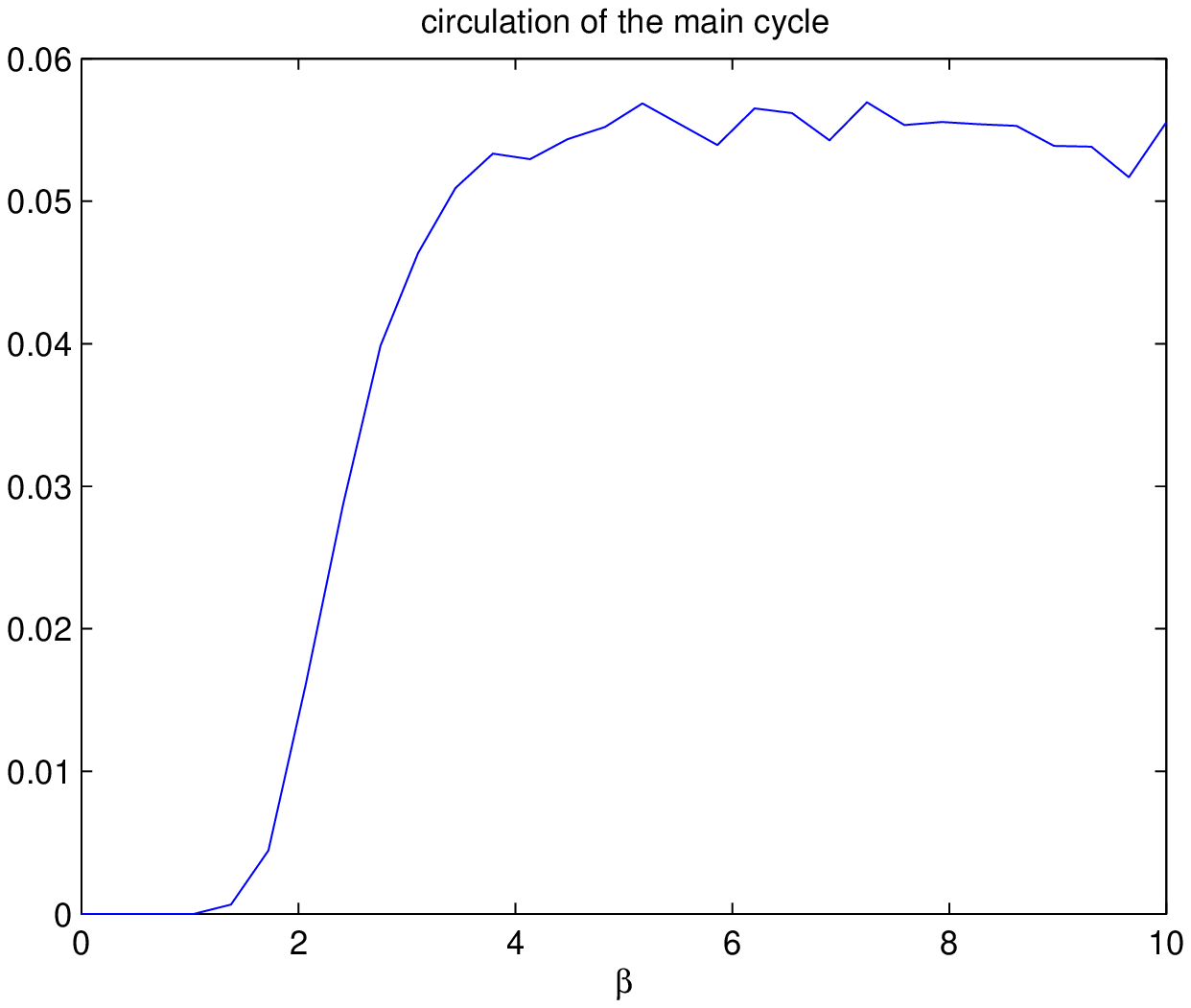}}
\caption[fig35]{Net circulation of the main cycle as a function of
the noise strength, $\beta$.}
\label{fig35}
\end{figure}

Finally, Fig. \ref{fig35} shows how the net circulation of the
dominant, main cycle varies with $\beta$. It is clearly seen that
the net circulation of the main cycle increases monotonically, which
implies the appearing of more and more distinct synchronization with
increasing $\beta$.  The direction of the net circulation does not
change.

The net circulations of all the cycles are very small when $\beta$,
$\alpha$ and $\gamma$ are near zero, since the system is close to
equilibrium state (reversible) according to lemma \ref{lem}.

And for large $\beta$, the net circulations of all the cycles except
the main cycle are very small, whose order are about $10^{-6}$ by
numerical simulation using the Monte Carlo method for circulation of
Markov chain according to Theorem \ref{Thm-circulation}. All the
circulations of non-main cycles actually decrease with increasing
$\beta$ when $\beta$ is large.

As in \cite{ZQ1}, we also notice that there exists an inflection
point in the curve in Fig. \ref{fig35}.  This implies a cooperative
transition of the net circulation of the main cycle while varying
the noise level $\beta$. We are currently studying the nature of
this phase-transition like behavior.  The implication of this
observation remains to be further elucidated.

It is also important to point out that the above results are very
robust with respect to the values of the matrix $T$ and the
threshold $U$, which is similar to the main result in
\cite{LLL04,ZQ1}.

The appearance of the peak in the power spectrum is a signature of a
coherent resonance; the existence of circulation means some
components of the system are synchronized. A mathematical
equivalence between the power spectral peaking and the circulation
was proved by Qian et.al. \cite{QQZ} in the case of Markov chains.
It has been generalized to all Markov processes by Jiang and Zhang
\cite{JZ}.

\section{Conclusion and Discussion}

Quantitative understanding of biological systems and functions from
their interacting components  presents a significant challenge as
well as an unique opportunity for scientists of diverse disciplines.
Based on the mathematical theory of nonequilibrium steady states,
especially the cycle circulation distribution, synchronized dynamics
in a stochastic yeast cell-cycle network is investigated in detail
and quantitatively, which is more definite and convincing than the
previous method often used by physicists and biologists.

{\bf Synchronization and circulation in stochastic network.}

As we know, the occurrence of a deterministic limit cycle in an
ordinary differential equation(ODE) model is the hallmark of a
synchronization phenomenon. For a stochastic system, such a definite
concept no longer holds and a logical generalization needs to be
investigated. In fact, the very meaning of synchronization requires
clarification. In this case, the concept of {\bf stochastic limit
cycle} is useful in describing the synchronization phenomenon. For
example, in our present model of yeast cell cycle, the trajectory
concentrates around a main cycle, which can be defined as a
stochastic limit cycle, is in many aspects very similar to the limit
cycle of a deterministic model. Although in many situations this
concept does not have a clear definition, it has been rigorously
discussed in the context of Markov chain under the heading of {\bf
circulation theory} for many years \cite{QQ82,QQQ84}. We recommend
the Chapters 1 and 2 of \cite{JQQ} for the systematic treatment of
this theory.

It is shown in the present paper that the stochastic synchronized
dynamics in the cell-cycle model, first observed in \cite{ZQ1}, can
be nicely characterized in the perspective of the theory of
nonequilibrium steady states and Markov chain circulations. The
stochastic Markov chain is reversible (equilibrium state) if and
only if the net circulation of every cycle vanishes, which is one of
the most important theorems in the mathematical theory of NESS.

The stochastic Boolean network is reversible if and only if the
matrix $T$ is symmetric, and the net NESS circulation emerges as
long as the probability of the directed cycle is larger than that of
its reversal, which is just the preconditions for synchronized
dynamics to occur. So circulation can be regarded as the
quantitative characterization of stochastic synchronized dynamics.
Those cycles with high circulation perform stronger synchronized
dynamics, while other cycles with low circulation perform weaker
synchronized dynamics. Moreover, the distinct global synchronized
dynamics can be obviously observed if there is a large scale
separation in the magnitude of the circulation between an unique
main cycle and the others.

On the other hand, circulation as a good characterization of
synchronized dynamics in stochastic models, corresponds perfectly to
the power-spectrum algorithm in determining the synchronized
dynamics. The magnitudes of the main cycle's circulation and the
dominant power-spectral peak both increases monotonically with
increasing $\beta$, showing more and more distinct synchronization
in the biological systems. Also the period of the dominant
power-spectral peak converges to the period(number of states) of the
main cycle with dominant circulations.

{\bf Stable attractor and robustness in the presence of biological
stochasticity. }

Biological systems have to be robust to function in complex (and
noisy) environments. More robustness could also mean being more
evolvable, and thus more likely to survive. In the model discussed
in the present paper, the unique limit cycle in the deterministic
network model and the monotonically increasing circulation in the
stochastic network model both demonstrate the stability of the main
stochastic limit cycle. It can be obviously seen that the
circulation of the main cycle dominate under a large range of noise
level (Fig. \ref{fig35}), which consequently means that this
biological function is very robust against small perturbations. It
is directly responsible for this cellular process.

Moreover, in some stochastic Boolean network model, one can contain
both stable and unstable stochastic limit cycles, where the unstable
stochastic limit cycle surprisingly causes vanishing circulation
when the noise intensity tends to zero, {\bf although it actually is
a true limit cycle of the corresponding deterministic Boolean
network model which attracts more than half the states in the state
space!} This phenomenon, called ``noise induced global attractive
behavior'', is recently observed by us in a stochastic Boolean
network model of the Siah-1/beta-catenin/p14/19 ARF loop of the
protein p53 pathways (unpublished work), in which the biological
trajectory, starting from the initial state where only p53 protein
is activated, is attracted to the unique stable stochastic limit
cycle of the protein states rather than the other unstable one. This
fact further confirms the robustness of this biological system.
Certainly, more implication of our observations remain to be further
elucidated in the future, when applied to other significant
bio-problems.

\section*{Acknowledgement}

The authors would like to thank Prof. Da-Quan Jiang in Peking
University for help suggestion. This work is partly supported by the
NSFC(Nos. 10701004, 10531070 and 10625101) and 973 Program
2006CB805900.


\small
\begin{thebibliography}{99}
\bibitem{Am} Amit, D.J.: {\em Modeling brain function: The world of attractor neural
networks.} Cambridge University Press 1989
\bibitem{tyson1}
Chen, K.C., Calzone, L., Csikasz-Nagy, A., Cross, F.R., Novak, B.,
Tyson, J.J.: Integrative analysis of cell cycle control in budding
yeast. {\em Mol. Biol. Cell} {\bf 15} 3841-3862 (2004)

\bibitem{tyson2}
Ciliberto, A., Novak, B., Tyson, J.J.: Mathematical model of the
morphogenesis checkpoint in budding yeast. {\em J. Cell Biol.}
{\em 163} 1243-1254 (2003)

\bibitem{FMWT} Fall, C.P., Marland, E.S., Wagner, J.M. and Tyson, J.J.: {\em Computational cell biology.}
 New York: Springer-Verlag  2002
\bibitem{Fox1} Fox, R.F. and Lu, Y.: Emergent collective behavior in large numbers of globally coupked independently stochastic ion channels.
 {\em Phys. Rev. E } {\bf 49}(4) 3421-3431  (1994)
\bibitem{Fox2} Fox, R.F.: Stochastic versions of the Hodgkin-Huxley equations.
{\em Biophys. J.} {\bf 72} 2068-2074  (1997)
\bibitem{Gi} Gillespie, D.T.:
   Exact stochastic simulation of coupled chemical reactions. {\em J. Phys. Chem.} {\bf 81}(25), 2340--2361 (1977)

\bibitem{Goldbeter}
Goldbeter, A.: {\it Biochemical Oscillations and Cellular Rhythms:
the Molecular Bases of Periodic and Chaotic Behaviour},
New York: Cambridge Univ. Press 1996

\bibitem{Ha} Hasegawa, H.: On the construction of a time-reversed Markoff process,
   {\em Prog. Theor. Phys.} {\bf 55}, 90--105 (1976); Variational principle for
   non-equilibrium states and the Onsager-Machlup formula, {\em ibid.} {\bf 56},
   44--60 (1976); Thermodynamic properties of non-equilibrium states subject to
   Fokker-Planck equations, {\em ibid.} {\bf 57}, 1523--1537 (1977);
   Variational approach in studies with Fokker-Planck equations, {\em ibid.} {\bf 58},
   128--146 (1977)

\bibitem{hill}
Hill, T.L.: {\em Free Energy Transduction and Biochemical Cycle Kinetics.}
New York: Springer-Verlag 1989

\bibitem{H82} Hopfield, J.J.: Neural networks and physical systems with emergent collective computational abilities.
{\em Proc. Natl. Acad. Sci. (79)}  2554-2558 (1982)
\bibitem{H84} Hopfield, J.J.: Neurons with graded response have collective computational properties like those of two-state neurons.
{\em Proc. Natl. Acad. Sci. (81)} 3088-3092 (1984)
\bibitem{H95} Hopfield, J.J., Herz, A.: Rapid local synchronization of action potentials: Toward computation with coupled integrate-and-fire neurons.
{\em Proc. Natl. Acad. Sci. (92)} 6655-6662 (1995)
\bibitem{JQQ} Jiang, D.Q., Qian, M. and Qian, M.P.:  {\em Mathematical theory of
   nonequilibrium steady states - On the frontier of probability and dynamical
   systems.}  (Lect. Notes Math. {\bf 1833}) Berlin:  Springer-Verlag 2004
\bibitem{JZ} Jiang, D.Q. and Zhang, F.X.:  The Green-Kubo formula and power spectrum of reversible Markoc processes.
   {\em J. Math. Phys.} {\bf 44}(10), 4681--4689 (2003)

\bibitem{Kei} Keizer, J.:  {\em Statistical thermodynamics of nonequilibrium processes.} New York:  Springer-Verlag
1987
\bibitem{SK} Kalpazidou, S.L.: {\em Cycle representations of Markov processes.} (Applications of Mathematics. {\bf 28}) Berlin: Springer-Verlag 1995

\bibitem{LLL04} Li, F., Long, T., Lu, Y., et al.: The yeast cell-cycle network is robustly designed.
{\em Proc. Natl. Acad. Sci. 101 (14)}  4781-4786 (2004)
\bibitem{Mc} McQuarrie, D.A.: Stochastic approach to chemical kinetics. {\em J.Appl.Prob.} {\bf 4}(3), 413-478 (1967)
\bibitem{Mu} Murray, J.D: {\em Mathematical biology, 3rd Ed.}
New York: Springer 2002
\bibitem{NP} Nicolis, G. and Prigogine, I.: {\em Self-organization in nonequilibrium
   systems: from dissipative structures to order through fluctuations.}
   New York: Wiley 1977

\bibitem{hqjpcm05}
Qian, H.: Cycle kinetics, steady-state thermodynamics and motors--a paradigm for
living matter physics.
{\em J. Phys. Cond. Matt.} {\bf 17}, S3783-S3794 (2005)

\bibitem{hqjpc06}
Qian, H.: Open-system nonequilibrium steady-state: statistical
thermodynamics, fluctuations and chemical oscillations.
{\em J. Phys. Chem. B.} {\bf 110} 15063-15074 (2006)

\bibitem{QH1} Qian, H., Saffarian S. and Elson E.L.:  Concentration fluctuation in a mesoscopic ocillating chemical reaction system.
   {\em Proc. Natl. Acad. Sci. USA} {\bf 99} 10376--10381 (2002)

\bibitem{hqprl05}
Qian, H. and Reluga, T.C.: Nonequilibrium thermodynamics and nonlinear
kinetics in a cellular signaling switch.
{\em Phys. Rev. Lett.} {\bf 94} 028101 (2005)

\bibitem{QG}
Qian, M.P. and Gong, G.L.: {\em Stochastic Processes(second
edition).} Peking University Press. 1997 (in Chinese)

\bibitem{QQ82} Qian, M.P. and Qian, M.:  Circulation for recurrent Markov chains.
   {\em Z. Wahrsch. Verw. Gebiete} {\bf 59}, 203--210 (1982)
\bibitem{QQQ84} Qian, M.P., Qian, C. and Qian, M.: Circulations of Markov chains
   with continuous time and the probability interpretation of some determinants.
   {\em Sci. Sinica (Series A)} {\bf 27}(5),  470--481 (1984)
\bibitem{QQZ} Qian, M, Qian, M.P. and Zhang, X.J.:  Fundamental facts concerning reversible master equations.
   {\em Phys. Lett} {\bf 309}, 371--376 (2003)
\bibitem{Sc} Schnakenberg, J.: Network theory of microscopic and macroscopic
   behaviour of master equation systems. {\em Rev. Modern Phys.} {\bf 48}(4),
   571--585 (1976)

\bibitem{Scott}
Scott, A.C.: {\em Neuroscience: A Mathematical Primer.}
New York: Springer-Verlag 2002.

\bibitem{sync}
Strogatz, S.: {\em Sync: The Emerging Science of Spontaneous Order.}
New York:Hyperion 2003

\bibitem{Van} Van Kampen N.G.:  {\em Stochastic Processes in Physics and Chemistry.}
 Amsterdam:North-Holland 1981
\bibitem{WD} Wilkinson, D.J.:  {\em Stochastic Modelling for Systems Biology.}
 Chapman and Hall/CRC 2006
 \bibitem{winfree}
Winfree, A.T.: {\em The Geometry of Biological Time,} 2nd. Ed.
Berline: Springer-Verlag 2000
\bibitem{YZ} Yan, P.F. and Zhang, C.S.: {\em Artificial neural networks and simulating evolutionary computation.} Tsinghua University Press
2005 (in Chinese)
\bibitem{Zhou} Zhou, T.S., Chen, L.N., and Wang, R.Q.: A mechanism of synchronization in interacting multi-cell genetic systems.
 {\em Physica D} {\bf 211}, 107--127 (2005)
\bibitem{ZQ1} Zhang, Y., Qian, M.P., Ouyang, Q., et al.: Stochastic model of yeast cell-cycle network.
 {\em Physica D} {\bf 219},  35--39 (2006)
\bibitem{ZQ2} Zhang, Y., Yu, H., Deng, M.H., Qian, M.P.: Nonequilibrium model for yeast cell cycle.
Computational Intelligence and Bioinformatics. International
Conference on Intelligent Computing, ICIC2006, Kunming, China,
August 16-19, Proceedings, Part III, 786-791 (2006)
\end{thebibliography}
\end{document}